# Web-based Visualization and Analytics of Petascale data: Equity as a Tide that Lifts All Boats

Aashish Panta*  Xuan Huang  Nina McCurdy  David Ellsworth  Amy Gooch
University of Utah  University of Utah  NASA Ames Research Center  NASA Ames Research Center  University of Utah

Giorgio Scorzelli  Hector Torres  Patrice Klein  Gustavo Ovando-Montejo  Valerio Pascucci
University of Utah  NASA Jet Propulsion Lab  Caltech  Utah State University, Blanding  University of Utah

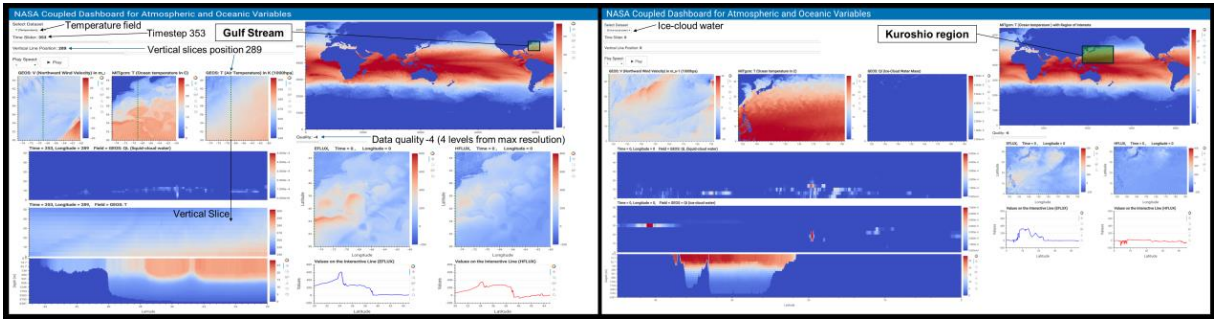

Figure 1: We provide unprecedented equitable access to massive data via our novel data fabric abstraction enabled by dashboards on commodity desktop computers with a simple weblink for everyone from top NASA scientists to students in disadvantaged communities to the general public. The two example dashboards for petascale climate data shown with multiple variables and 10,000-time steps: the Gulf Stream region dashboard on the left and the Kuroshio region dashboard on the right.

**ABSTRACT**

Scientists generate petabytes of data daily to help uncover environmental trends or behaviors that are hard to predict. For example, understanding climate simulations based on the long-term average of temperature, precipitation, and other environmental variables is essential to predicting and establishing root causes of future undesirable scenarios and assessing possible mitigation strategies. While supercomputer centers provide a powerful infrastructure for generating petabytes of simulation output, accessing and analyzing these datasets interactively remains challenging on multiple fronts. This paper presents an approach to managing, visualizing, and analyzing petabytes of data within a browser on equipment ranging from the top NASA supercomputer to commodity hardware like a laptop. Our novel *data fabric abstraction* layer allows user-friendly querying of scientific information while hiding the complexities of dealing with file systems or cloud services. We also optimize network utilization while streaming from petascale repositories through state-of-the-art progressive compression algorithms. Based on this abstraction, we provide customizable dashboards that can be accessed from any device with any internet connection, enabling interactive visual analysis of vast amounts of data to a wide range of users - from top scientists with access to leadership-class computing environments to undergraduate students of disadvantaged backgrounds from minority-serving institutions. We focus on NASA's use of petascale climate datasets as an example of particular societal impact and, therefore, a case where achieving equity in science participation is critical. We validate our approach by improving the ability of climate scientists to visually explore their data via two fully interactive dashboards. We further validate our approach by deploying the dashboards and simplified training materials in the classroom at a minority-serving institution. These dashboards, released in simplified form to the general public, contribute significantly to a broader push to democratize the access and use of climate data.

## 1 INTRODUCTION

Despite the daily generation of massive petabyte-scale datasets, effective analysis and visualization remain critical for extracting valuable insights. Scientific institutions like NASA are at the forefront of generating such data. While significant resources are poured into making it accessible (e.g. NASA's DYAMOND and ECCO data [42, 41]), inherent challenges persist. However, these datasets present unique challenges: difficulty accessing the data, limitations in computational power, and the need for real-time processing capabilities. Downloading petascale data locally is problematic due to limitations of local memory or disk size and insufficient bandwidth for remote disks [13]. Our work focuses on collaboration with these institutions to improve data accessibility, empowering researchers to unlock the hidden knowledge within these vast datasets.

Researchers and scientists often want to ask conceptually simple questions, such as viewing time-series data and slices of volumes. Petascale data visualization can require up to hundreds of GPU and CPU core hours, which can take hours of waiting in a queue at a busy center. Researchers also may need to interactively visualize large datasets, which is difficult with traditional static visualization methods that limit the ability to ask real-time questions and perform on-the-fly analysis. Other technical challenges for big data in domains like climate science include migrating code, analytic products, and large repositories within the growing network of storage and computational resources [16].

Commonly, static visualizations are generated for a selected time range, scalar, region, and resolution, limiting the ability to interactively view and analyze the data. Our collaboration with domain scientist and visualization experts has helped to create a novel in-teractive visualization dashboard for petabytes of data with progressive loading and decoupling of the storage infrastructure in order to increase data democratization. Our specific contributions include:

- A novel *data fabric abstraction* layer that allows users to request information based on their scientific needs without dealing with the low-level specification of file formats or network speed. Our *FAIR Digital Objects (FDOs)* layer responds to user requests within the specified quality/resource bounds or notifies that the request may need to be revised (e.g., reducing quality or increasing resources). Our approach allows undergraduate students in a minority-serving institution to use in their exercises the same petascale dataset as NASA scientists on their largest supercomputer, seeking to advance data access equity in science.
- Efficient Data Reorganization, Conversion, Reduction/Optimization pipeline that allows for efficient storage and data transfer by utilizing compression strategies for data and transforms the data into an *Analysis-Ready Cloud-Optimized (ARCO)* friendly format to significantly reduce the computational load and storage requirements.
- Scalable Visualization Dashboards enable progressive visualizations of petascale data with advanced analytical tools and a user-friendly design, encouraging scientific curiosity and discovery.
- Data Democratization via publicly accessible web links to more than a petabyte of data on the cloud in an optimized format, enhancing public access and collaborative opportunities.

---
*e-mail: aashishp@sci.utah.edu

We demonstrate our approach through dashboards available on supercomputers and servers, both accessing publicly available data from cloud storage. The phrase in the title, "Equity as a Tide that Lifts All Boats," is a nod to the proverb "a rising tide lifts all boats", which dates back to the early 1900s. The proverb is typically used in economics and is used here about the benefits of providing data access equity. We evaluate our dashboards through three use cases: 1) petascale visualization of multiple variables with 1.1 PB data from the cloud; 2) studying relationships between oceanic and atmospheric variables to see how the change in oceanic phenomena such as sea surface temperature affects the formation of ice, water, and clouds in the atmosphere using cloud-served data; and 3) demonstrating how the "rising of the tide" to data equity opens the doors for underserved communities to access data previously out of reach. We also examine performance, discuss lessons learned, and review the intellectual merit and societal benefits of our novel approach for petascale data visualization and analysis.

## 2 RELATED WORKS

Visualizing large-scale data directly from a web environment provides unprecedented access to information. The ability to process and render complex datasets from a web browser offers unique advantages in efficient analysis, accessibility, and data management. The shift toward browser-based visualization tools enables users across various disciplines to access and interact with information in real-time without the need for specialized hardware or extensive software installations. As the interest and demand for web-based visualization grows, researchers and developers have developed a variety of frameworks, libraries, and methodologies to tackle the inherent challenges associated with rendering large-scale datasets in a browser environment.

### 2.1 Large-Scale Web-based Visualizations

The most popular web-based visualization libraries today include D3 [11], popular for its ability to directly manipulate and transform the contents within its *document object model (DOM)*. Several other libraries leverage WebGL to handle large-scale data visualizations efficiently and with high performance directly within a browser. WebGL enables these libraries to provide rich, interactive, and 3D visualizations using the GPU for graphics rendering. Among these are libraries Deck.gl [70], Luma.gl and Three.js [5].

Visualizing large-scale datasets from a browser has been previously researched. Usher et al. [65, 63] developed an isosurface computation algorithm for block-compressed data to visualize a terabyte of scientific data from a browser. We have far exceeded this by creating a framework that can visualize more than a petabyte of data from the cloud. Alder et al. [4] developed USGS National Climate Change Viewer to visualize 17 terabytes of climate projection data from compressed NetCDF-4 files and preprocessed the data for statistics instead of computing them on the fly. Walker et al. [69] worked on 50 megabytes of geospatial data, mentioning a browser limitation that caused excessive latency. Other tools, such as ParaviewWeb [28], perform data processing and rendering on the server side and stream back the results to the client. Mohammad et al. [51] deployed efficient and affordable scientific visualization as a microservice, but noted challenges with network inconvenient latency and costly egress costs. However, none of these systems have been capable of working with petascale datasets.

A framework developed by Lu et al. [36] allows on-the-fly visualization of the multiscale climate datasets but remotely uses cloud services to provide big data processing ability. Ravindu [50] describes loading data into memory as one of the key challenges of visualizing data from a browser. To tackle these challenges, tools such as Firefly [21] use progressive rendering techniques on a dataset of size 250GB to visualize from a browser. Other challenges mentioned by Khadija et al. [8] include scalability and high-performance requirements for big data analysis and visualization.

### 2.2 Climate-specific Data Visualization Tool

Some existing tools, such as Ultrascale Visualization Climate Data Analysis Tools (UV-CDAT) [72], support data gridding and exploratory data analysis but require users to download sophisticated packages. A Python-based tool called CCPviz developed by Aizenman et al. [3] provides a data processing and visualization module for climate data. However, the CCPviz architecture is inefficient for petascale data as it requires transferring all selected data between the different layers of their architectures. Sun et al. [61] developed a web-based visualization framework for climate data using Google Earth by precomputing all the images at the local server and mentioned that on-the-fly data access from remote servers is slow and impractical.

A web analysis platform called ClimateCharts.net developed by Zepner et al. [77] focuses on the general interactive features of the data but had issues due to the lack of computational resources and network latency. Another framework developed by Wong et al. [74] provided scalar fields and flow visualization but required significant data downsampling to make their workflow run on the desktop computer. Other challenges mentioned by Wong et al. [73] include being dependent on in-situ analysis of data, lack of interaction, visualization techniques, and limited community engagement.

Several implementations endeavor to make the NASA Climate data easier to access [18, 1, 7, 23]. Such methods require users to access libraries designed to make the data accessible. Still, the ability to slice vertically or horizontally interactively requires manual installation of complex libraries or additional expertise. Scientists and developers have created libraries like xmitgcm [75, 2] to handle these issues for petascale NASA datasets, but they still lack the interactive features that many climate scientists desire. Other tools developed at NASA, such as Podaac [40] provide *near real-time (NRT)* access to some data products but still do not give users the flexibility to change the range, colormap, and other utilities such as slicing the data, helpful in navigating the depths of the data. Ellsworth et al. [15] developed an environment to visualize these petabytes of data using the hyperwall, a display wall with 128 displays, and associated computing clusters. Whereas the high-resolution displays and custom software allowed for quickly viewing large amounts of data, this unique system was restricted to scientists who could secure an invitation to the facility.

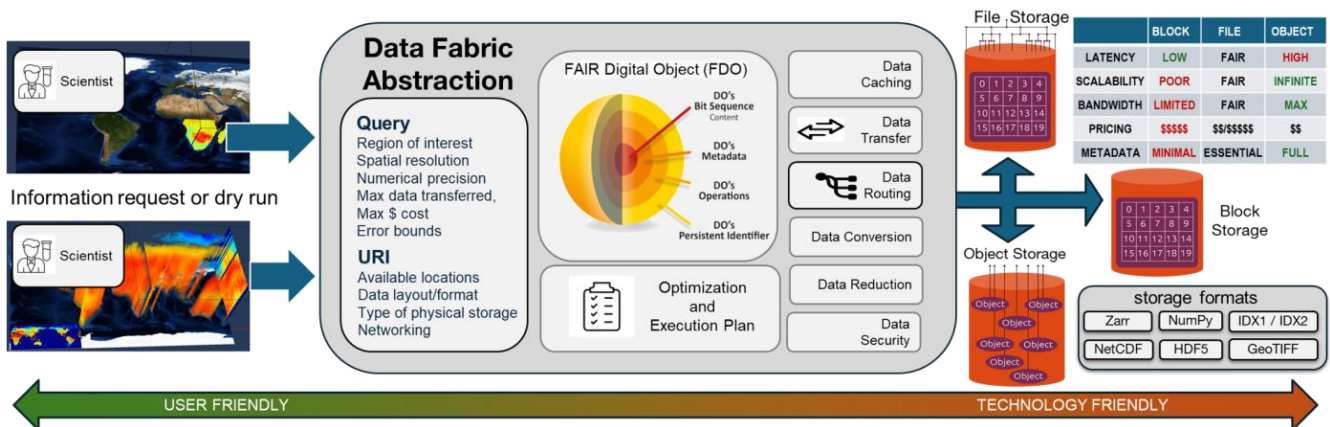

Figure 2: Our novel data abstraction framework allows a scientist to express a query for the information needed with additional parameters, such as the quality required to achieve a trustworthy result and/or the maximum amount of resources available for its execution. Our Data Fabric Abstraction (left) handles the query and builds the *Uniform Resource Identifier (URI)*. The *FAIR Digital Object (FDO)* provides the information necessary for an implementation that optimizes the execution of the query (middle). The low-level execution (right) will use the available networking and storage resources, including different file formats and storage models (file systems, object storage, or block storage), as needed.

## 2.3 Data Reorganization using OpenVisus

Many scientists need help dealing with massive datasets due to hardware limitations, slow data movement, and I/O bottlenecks. Among other technologies, we use OpenVisus [47, 44], an open-source out-of-core data management framework to reorganize the data. The framework employs multiresolution space-filling curves for data reorganization and interactive exploration [60]. OpenVisus supports fast I/O of petascale simulations [32, 31] as well as post-hoc querying and visualizing petascale data in various scientific applications [64, 20, 47, 49]. Designed to provide progressive random access for very large datasets, OpenVisus optimally exploits the existing caching hardware in modern architectures. The cache-oblivious approach [67, 46, 76] exploits this structure by storing large data arrays in a cache-optimized manner. A critical aspect that will be specialized and optimized, especially for use cases presented in this paper, is the ability to progressively encode spatial resolution and numerical precision of the data [24, 9], thereby minimizing the cost of data movements for any data analysis and visualization workflow [22].

## 3 A NOVEL DATA FABRIC ABSTRACTION

We introduce a data abstraction layer that concurrently addresses the user's need to access information easily while being able to control the amount of resources used. Via the use cases presented in this paper, we demonstrate how the data abstraction layer aids in visualizing, analyzing, and sharing petascale climate simulation datasets. We show how our framework facilitates using these massive datasets for world-renowned climate scientists using NASA's largest supercomputer and for undergraduate students in a minority-serving institution, mostly native Americans from the Navajo Nation (typically first-generation college students).

Our data fabric abstraction consists of several modules as shown in Figure 2 including query, universal resource identifier (URI), fair digital object (FDO), and plan modules, as well as pipelines for data caching, transfer, routing, conversion, reduction, and security. The data fabric abstraction then relies on an API to access either file, block, or object storage. The storage formats include ZARR, numpy, NetCDF, HDF5, GeoTIFF, and IDX1/IDX2. By building this modular abstraction, we can easily add, change, and update any of these black boxes through a simple API without worrying about cascading issues.

## 3.1 Query and URI

To alleviate users from the complexities associated with low-level storage intricacies, we provide an API to request information at a high level of abstraction and include a variety of technical requirements. Typical primary query elements include spatial extent, time value or range, and variables of interest (temperature, salinity, velocity, etc.). Queries are specified not with respect to how the data is stored in a particular file format but in the analysis coordinate system, similarly to an Xarray API [26]. While Xarray is user-friendly, Xarray does not address the problem of mapping a request to an impractically large amount of resources. We, therefore, introduce additional parameters that the scientific community has identified as needed when dealing with massive data [29, 12, 6]. For example, for a given query, the user can specify the spatial resolution and/or numerical precision needed to satisfy the scientific needs. Additional constraints include the maximum cost in egress fees budgeted for data stored in the commercial cloud or the maximum delay between query and response. Since it may not always be possible to satisfy all the requirements, a query may not return the information requested but indicate that the conditions need to be relaxed (Figure 2, left).

As users make specific requests within the query abstraction framework, the back-end *uniform resource identifier (URI)* abstraction handles query requests, such as data caching, transfer, routing, conversion, reduction, and security. Each of these is critical for maintaining performance and data integrity across different storage types and formats, whether in block storage, file storage, or object cloud storage. Flexible data management strategies can optimize for I/O, cost, speed, or accessibility, depending on the user's needs.

## 3.2 FAIR Digital Object

To address the complexities of large data distributed across various platforms and the potential use of different file formats, we introduce the first advanced *FAIR Digital Object (FDOs)* framework [58] implementation, which democratizes access to data while following the *FAIR (Findable, Accessible, Interoperable, Reusable)* guiding principle [71], illustrated in Figure 2. In particular, an FDO generalizes the concept of *Digital Object Identifier (DOI)* with the inclusion of executable elements such as a bit sequence, operations, metadata, and an identifier. Practically, the FDO includes all the actionable information needed to determine if and how a given query can be resolved (Figure 2, middle).

### 3.3 Storage and File Formats

The last module of the data fabric abstraction includes the API between the data and the storage. We capitalize on OpenVisus's capabilities and improve its use as a cloud or caching data model for fast data processing. In particular, we tackle the challenge of working with datasets too large for a system's memory by utilizing OpenVisus' out-of-core computations.

We currently have modules for seamlessly exploiting file, block, object, and distributed storage options. File storage functions as network-attached storage. Block storage offers high I/O performance, strong consistency, and low-latency connectivity. Object storage guarantees high availability and is durable and infinitely scalable on the cloud. Distributed storage [27, 14, 30] is tailored for long-term scientific research and is immutable, verifiable, and cost-effective through incentive systems, smart contracts, and quality of service tradeoffs [33, 66, 55] (Figure 2, right).

Our abstraction layers allow transparent data conversion between many file formats and optimize storage and retrieval without user intervention. For example, a query for a high-resolution climate model might be stored in Zarr format in the cloud but could be automatically converted to a more compact representation for the user, like GeoTIFF or NetCDF, if that is what they require for their analysis.

### 3.4 Transforming HPC Data for Cloud Storage

The workflow illustrated in Figure 3 provides an overview of converting High-Performance Computing (HPC) data to cloud data. The process begins with a simulation model running on the HPC system. This model generates raw data at a petascale level that requires substantial storage and processing power. The first step converts the generated raw data to Analysis-Ready, Cloud Optimized [2, 56] IDX format. The OpenVisus IDX format enhances our ability to process large-scale datasets at unprecedented speeds. Our conversion pipeline achieves remarkable efficiency, converting data at 1 TB per hour using only 16 cores. This rate suggests that, with 16,000 cores, it would be possible to reach conversion speeds of up to 1 PB per hour, depending on the system's I/O capacity. The conversion to IDX format is motivated by the need for a more flexible and performance-optimized approach to handle large-scale data. Traditional formats often struggle with the demands of petascale datasets, resulting in extended periods of data preparation and subsequent delays in analysis. In contrast, the optimized IDX format works with devices with minimal resources and high-performance computing environments. The IDX format enables significantly more efficient operations for reading and writing data, facilitating faster data manipulation and visualization.

Adding an ARCO-friendly format further streamlines the data, significantly reducing computational load and storage requirements and allowing efficient and direct access to data subsets in the cloud. Afterward, data compression techniques such as ZIP, Zlib, and LZ4 for lossless compression and ZFP for lossy compression reduce the size of the data, which helps speed up the transfer process and lowers storage costs. The compressed IDX dataset is then securely transferred to the cloud via 'secure copy protocol.' These efficient and secure data transfer protocols synchronously move large volumes of data from the HPC environment to cloud storage. In addition, our frameworks also manage an IDX local cache to optimize data queries, ensuring faster access and improved performance for end-users accessing the data from notebooks and dashboards.

In our empirical evaluation, we found a significant influence of block sizes on the amortization and prediction of network latencies and the achievement of optimal bandwidth transfers. By examining various block sizes ranging from 32KB to 16MB, we investigated their impact on the overall workflow execution; altering block sizes achieves an immediate effect on the total number of generated objects, which could swiftly escalate to hundreds of millions, potentially leading to operational delays in any network file system.

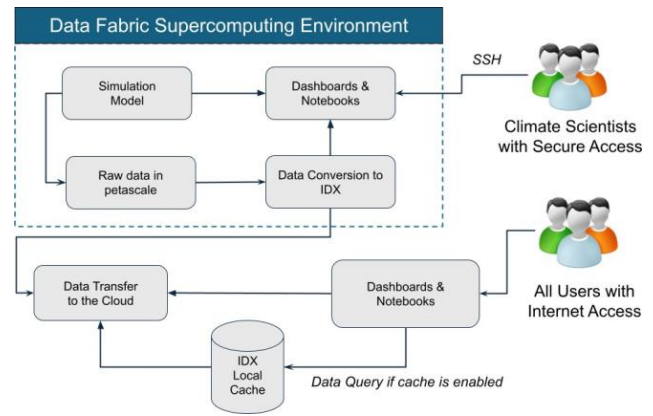

Figure 3: An example workflow showing data conversion and retrieval pipeline for Jupyter notebook and dashboards from a supercomputing environment.

Through practical experimentation, we established that an optimal block size, identified between 1MB and 8MB, closely aligns with those utilized by leading commercial entities in their File Sharing Solutions (e.g., Google Drive, Dropbox, Microsoft One Drive), particularly when coupled with effective client-side caching.

### 3.5 Results of DFA: Decoupling Data From Storage

Decoupling data from its storage infrastructure is paramount for achieving longitudinal data access and sharing capabilities. This separation is crucial because the lifespan of any physical storage medium is inherently shorter than that of the data it holds. Given the rapid evolution of technology and business landscapes, today's optimal storage solutions can swiftly become prohibitively expensive or outdated. Therefore, data repositories must employ technology-agnostic abstractions that facilitate hybrid usage and seamless migration, minimizing costs and disruptions to user access.

Our visualization framework seamlessly supports: 1) Reading data at varying resolutions within Regions of Interest (ROIs), limiting the result to the available memory or the maximum number of projected pixels on the screen; 2) Generating summary videos from temporal data with specific resolution constraints; the total number of frames is established depending on the network bandwidth and the existence of pre-cached data; 3) Writing multiple versions of data, one for archival at low cost and one resolution-capped for quick sharing purposes.

Furthermore, user requests can be translated to different encoding and compression schemes depending upon several factors, e.g., high-but-slow compression for less frequently accessed data and low-but-fast compression for frequently accessed data. Additionally, our framework facilitates data migration between different storage tiers (hot, warm, and cold) [48] and enables transparent rerouting of data requests from local storage to external storage as needed. To make datasets publicly accessible, we upload petabytes of data after compression to Seal Storage, an S3-API-compatible decentralized cloud storage service [54].

### 3.6 Impact of our Data Fabric Abstraction

Our progressive streaming ability, combined with the cloud-served data in analysis-ready format, allows user access and visualization of large datasets without downloading the entire file or region. Our frameworks enable convenient remote collaboration and data access with standalone Jupyter notebooks [52] and dashboards via simple lines of code shown in Figure 4.

## 4 DASHBOARDS

We have found that whereas generating a simulation for a given data set and parameters may require hundreds of hours or more on a supercomputer, viewing that data should no longer require heavy

```python
import OpenVisus as ov
endpoint="https://maritime.sealstorage.io/api/v0/s3"
url= f"{endpoint}/utah/nasa/" \
     f"dyamond/mit_output/llc2160_arco/visus.idx"\
     f"?access_key=any&secret_key=any" \
     f"&endpoint_url={endpoint}&cached=arco"
db=LoadDataset(url)
data=db.read(x=[x_min,x_max],
             y=[y_min,y_max],
             z=[z_min,z_max])
```

Figure 4: Simple Python code fragment for accessing a petascale data stored on the Seal Storage cloud. The result of an input URL given to the **LoadDataset** function is assigned to **db**. The **db.read** returns a NumPy array that can easily be used in Python or Jupyter Notebooks.

computation. To solve this significant gap between the fast reading and visualization of massive data, we have integrated a Python version of OpenVisus [47] called OpenVisusPy [45] with web-based visualization frameworks such as Bokeh [10] and Panel [25]. These flexible and interactive widgets support a wide range of visualization techniques, allowing users to dynamically explore, analyze, and understand massive datasets with ease. An integrated environment enables users to interact with their data in previously impractical or too resource-intensive ways, allowing data-intensive analysis to become more accessible and insightful.

Our dashboard framework, an example shown in Figure 5 and in the supplemental materials, provides a diverse array of features designed to accommodate both casual explorers and scientific researchers, including dataset selection, region of interest extraction, timestep slider, horizontal and vertical slices, color map/palette, colormap range (user or dynamic), resolution sliders, playback functionality, and time speed control.

Our dashboard represents a significant advancement in the visualization and analysis of large-scale data, not limited by the size of the data, disk space, and available memory. It allows multiple interactive windows that show streaming progressively loaded slices of volume data, graphs of pixel values through the volume, or macro views of the dataset, as shown in Figures 1 and 5. By providing interactive tools and features, we aim to make complex data visualizations accessible and insightful for a broad audience, from researchers and scientists to educators and policymakers.

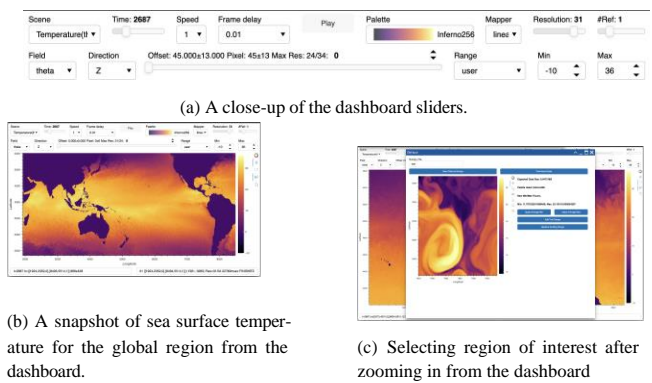

(a) A close-up of the dashboard sliders.

(b) A snapshot of sea surface temperature for the global region from the dashboard.

(c) Selecting region of interest after zooming in from the dashboard

Figure 5: Interactive visualization of sea-surface temperature for Use Case 1 and inset for selecting the regions of interest (left). The dashboard provides the ability to directly download the data locally or to download a Python script that fetches the region from a cloud (right).

## 5 EXAMPLES

Although our framework can work for many petascale gridded datasets, we demonstrate our framework for two large-scale climate simulation datasets: 1) the NASA 1.8 PB DYAMOND dataset [57, 38] and 2) the LLC4320 Ocean dataset [37, 18]. The Coupled Ocean-Atmosphere Simulation (COAS) run at *NASA Advanced Supercomputing (NAS)* is part of the international project called "Dynamics of the Atmospheric general circulation Modeled On Non-hydrostatic Domains" or DYAMOND. The purpose of COAS is to better understand the oceanic and atmospheric mechanisms that link air-sea interactions with the Earth's water cycle and extreme atmospheric events. As illustrated in Figure 1 and shown in the supplementary videos, our dashboards help solve the challenge of putting together all the data, providing access to efficient visualizations in 3D space of multiple atmospheric and oceanic variables. In this section, we will give an overview of the datasets, describe the specific details related to converting the data, especially the data compression gains, and discuss the performance of reading these data from different locations, such as NAS native file systems, locally cached cloud storage, and uncached cloud storage.

### 5.1 Dataset Overview

The first dataset, DYAMOND [57, 38], is the simulation output from research into coupling two models: a global atmospheric model and a global ocean model that were originally designed to be run separately. The atmospheric model is a C1440 configuration of the Goddard Earth Observing System (GEOS) atmospheric model running on a cubed-sphere grid. The global ocean model is an LLC2160 configuration of the MITgcm model that uses a lat-lon-cap grid. Each model was run for over 10000 hourly timesteps covering over 14 simulation months. The atmospheric model output has 20 3D and over 100 2D scalar fields, and the ocean model output has 5 3D and 15 2D fields. Both models have 3D fields such as temperature, north-south velocity, and east-west velocity. Some example fields from the atmospheric model are humidity, soil wetness, snow cover, and various cloud state variables. The ocean model includes fields of salinity, sea ice thickness, and freshwater flux, with a total size of approximately 1.8 petabytes.

Another ocean dataset called LLC4320 [37, 18] from the project called 'Estimating the Circulation and Climate of the Ocean', also known as ECCO, is the product of a 14-month simulation of ocean circulation and dynamics using MITgcm (MIT General Circulation Model). This simulation is very similar to the ocean portion of the DYAMOND coupled simulation, but was run with half the horizontal grid spacing (four times the cell count), and with values at the ocean surface boundary that were derived from observations and physical models. The model output has 5 3D and 13 2D scalar fields that include temperature, salinity, the three velocity components, sea ice, and radiation. This massive dataset is 2.8 PB in size.

We have built several interactive dashboards to potentially improve our understanding of global ocean circulation and its role in Earth's climate system. We provide the ability to integrate the DYAMOND and LLC4320 datasets as well as provide the large-scale simulation datasets stored on the cloud and on the NAS Pleiades Supercomputer [39], as shown in Figures 1, 5, 6, and 7. The instructions to access the existing deployed dashboards or launch a new one can be accessed from our GitHub repository [53]. We have worked on several GEOS and MITgcm simulation fields, converting them to IDX format, enabling seamless visualization and interaction via Jupyter notebooks and dashboards without intensive computational resources. We also collaborate with several scientists from NASA Jet Propulsion Lab and NASA Ames Research Center to help facilitate the extraction of their region of interest, especially the Gulf Stream and Kuroshio regions (Figure 1), and have built unique dashboards that display coupled outputs from both GEOS5 and MITgcm configurations of the simulation. Thus, our dashboards combine multiple petascale datasets into a single interface, allowing unprecedented visualization, interaction, and analysis of the sheer volume of data. This integration facilitates a deeper understanding of complex climatic phenomena by enabling scientists to seamlessly navigate and explore data across various scales and dimensions. Sections 5.3.1 and 5.3.2 will discuss specific use cases with the DYAMOND and LLC4320 datasets.

| NASA Dataset Model/Configuration | Original Size | Compression Algorithm | Precision | Reduced Size | Compression Factor |
|---|---|---|---|---|---|
| DYAMOND-GEOS/C1440 | 250 TB | ZIP | Lossless | 65.9 TB | 3.79 |
| | | ZFP | 30 bits | 152 TB | 1.65 |
| | | ZFP | 16 bits | 64.8 TB | 3.85 |
| ECCO-MITgcm/LLC4320 | 400 TB | ZFP | 16 bits | 112 TB | 3.57 |
| DYAMOND-MITgcm/LLC2160 | 1.1 PB | ZIP | Lossless | 440 TB | 2.49 |
| | | ZFP | 16 bits | 138 TB | 8 |

Table 1: Lossy and lossless compression algorithms on simulation dataest with final size and resulting compression factor.

## 5.2 Data Compression Examples

After converting datasets to IDX, and before we upload the data to the cloud, we apply compression to the datasets to reduce file size without compromising data. We have extensively evaluated various compression algorithms on the NASA datasets, using both lossy (with several precision bits) and lossless compression algorithms. The results achieved with compression are significant, as shown in Table 1. Based on initial tests, we found ZIP and ZFP to be the best compression algorithms for our data. We report on our examination of the lossless ZIP compression algorithm and the lossy ZFP algorithm [34] with 16-bit or 30-bit precision on a range of sampled regions of climate datasets.

A 250TB portion of the atmospheric data was first encoded to the streaming format and then processed through the lossless default-level ZIP compression algorithm to reduce its size to around 65.9 TB with a compression factor of 3.79. Table 1 shows compression rates for the same dataset under ZFP 30-bit and 16-bit compression.

We performed the ZFP compression at the 16-bit precision level for another 400TB of a compressed version of the LLC4320 dataset already available in the Pleiades [39] and brought this down to 112 TB with a compression factor of 3.57. Based on our preliminary tests, we decided to use default-level lossless ZIP compression for our use cases. For all oceanic fields included in the 1.1 PB dataset, the final size after compression was 440TB with a compression factor of 2.59. Although ZFP offered a slightly higher compression ratio, the data loss was significant for some of our collaborators.

## 5.3 Application-Specific Dashboards

Collaborating with researchers at NASA Ames Research Center and NASA Jet Propulsion Lab, we built several application-specific dashboards to demonstrate our framework. The first two use cases below describe a brief user study and their feedback. We also provide a use case with our collaborator at Utah State at Blanding, Native American Serving Non-Tribal Institutions (NASNTI).

### 5.3.1 Use Case 1: Multivariate Petascale visualization

To produce useful interactive analysis on massive datasets such as the NASA DYAMOND or LLC4320 Ocean Dataset, visualization scientists typically need to use computing resources at the NASA Advanced Supercomputing (NAS) facility. Although NAS supports and promotes full and open data access to the public, analyzing the data on the supercomputers requires logging into secure platforms and requesting nodes/cores. This dramatically limits the people who may be able to use the data in practice. Scientists and researchers might need to wait hours to days to load the data and produce a video clip for climate scientists, who then perform their scientific tasks on these fixed-resolution animations. Because supercomputing centers often store only full-resolution simulation data, accessing full-domain, full-resolution simulation data can take time, making quick turnaround on-the-fly analysis complex or slow.

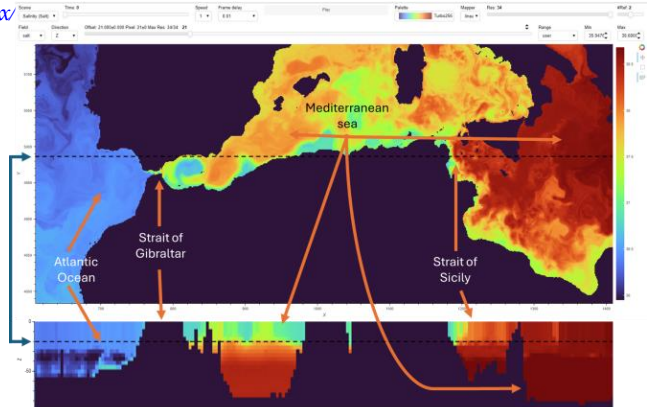

Figure 6: Case 1: Zoomed-in view of the general water circulation through the Strait of Gibraltar connecting the Mediterranean with the Atlantic Ocean.

However, our innovative dashboard and analytical approach significantly streamline working with massive datasets, such as those often encountered in climate science. We reduce the need for extensive computing resources, allowing the analysis to be performed on more accessible platforms without compromising accuracy. Users can bypass the inconvenient process of logging into secure platforms and waiting for supercomputer access. Instead, they can directly engage with the data through our user-friendly interface that offers real-time analysis and visualization capabilities.

One key objective of the dashboard shown in Figure 5 is to enable the visualization of multiple oceanic variables over time. Traditional visualization techniques struggle with the scale and complexity of the datasets involved, particularly when dealing with simulation outputs spanning 10,000 timesteps across multiple fields. Our dashboard addresses this challenge by offering progressive visualization capabilities and allows scientists to explore the data seamlessly through an intuitive interface. For example, Figure 6 shows the interesting phenomenon of water circulation around the Mediterranean region. The less saline water from the Atlantic Ocean passes through the Strait of Gibraltar and starts moving eastward. As the water moves east and the evaporation continues, the salinity tends to increase and starts sinking. Oceanographers and climate scientists worldwide have studied this interesting phenomenon, but no tool has ever allowed its interactive illustration on real data to the general public until now. Another example is around the Agulhas region, as shown in Figure 7, where the warm water from the Agulhas current flows along the southeast coast of South Africa and encounters the colder Atlantic Ocean, thus leading the current to bend back on itself. This process, also known as "retroflection," leads to the formation of large swirling masses of water, creating the Agulhas rings [43]. The dashboard enables any user to examine and interactively explore any regions of interest, as well as play the data across time without being restricted to time waiting for animations to render.

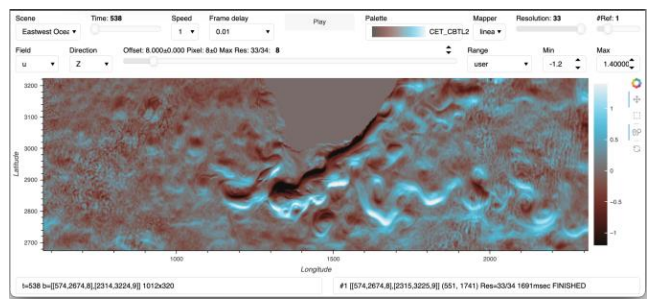

Figure 7: Formation of Agulhas rings at the African southeast coast demonstrated using the LLC2160 ocean dataset.

### 5.3.2 Use case 2: Oceanic and atmospheric variables

Our second use case builds on an existing collaboration between visualization researchers at NASA Ames Research Center and ocean scientists at JPL/Caltech. The collaboration sought to further investigate the impact of mesoscale and submesoscale ($< 500$ km) sea-surface temperature anomalies on local atmospheric circulation and vice versa. A frame from the animated version of figures from the published results of Strobach et al. [59] is shown in Figure 8.

Ocean-atmosphere interactions have long been considered to be limited to only the *atmospheric planetary boundary layer (APBL)*, up to 2000 m above the surface) The surface ocean is a mixed layer (50 to 200 m deep), with the atmosphere forcing the ocean. For example, strong winds deepen the ocean mixed layer, leading to a decrease in the *sea surface temperature (SST)* [19].

Numerical ocean-atmosphere models coupled with increasing spatial resolutions are challenging previously held theories. It is now known that the ocean-forced variability in the atmosphere is at scales smaller than 500 km, similar to the scales of ocean mesoscale eddies, which constitute up to 80% of the total ocean kinetic energy [19]. Turbulent heat and humidity fluxes are strongly enhanced above warm mesoscale eddies where convection develops and is reduced over cold eddies, leading to a significant net heating and humidification of the atmosphere. In addition, this impact of ocean eddies is not confined to the APBL but concerns the whole troposphere, which is up to 12,000 m above sea level. Through these mechanisms, the heat and humidity fluxes associated with ocean eddies intensify atmospheric storms traveling eastward [17] as shown in Figure 9. As a result, ocean eddies in the Kuroshio-Extension region off Japan can increase precipitations over the West Coast of the U.S.A. by 20% [35]. These results highlight the impact of ocean eddies on the Earth's water cycle and extreme atmospheric events. Recent studies [59] point to the important role of sea surface temperature (SST) fronts (10 km wide) surrounding ocean eddies on ocean-atmosphere exchanges: SST fronts trigger a secondary circulation, with the same width, in the atmosphere above the APBL that carries heat and humidity to the upper levels.

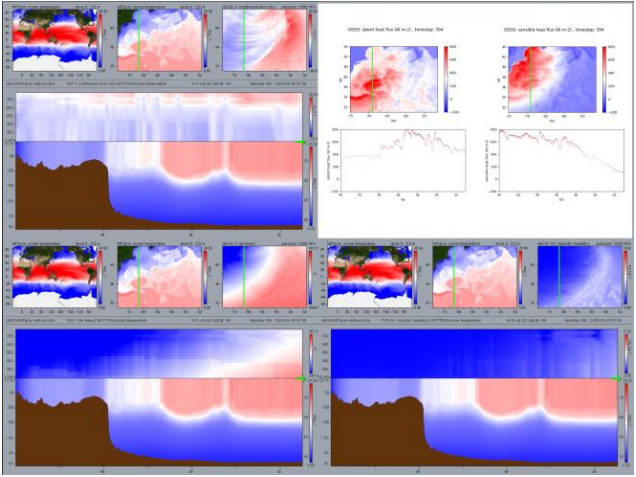

Figure 8: Prior preliminary animation by Nina McCurdy at NASA Ames Research Center created before our collaboration that motivated the creation of the dashboard shown in Figure 1. Image and video courtesy of Nina McCurdy, Copyright NASA 2023.

Over a 6-week period of intermittent collaboration and iteration in Spring 2023, the visualization researchers at NASA Ames Research Center and the ocean scientists at JPL/Caltech developed a preliminary visualization showing coupled vertical and horizontal slices of various fields of interest (ocean temperature, air temperature, northward wind velocity, specific humidity, latent heat flux, and sensible heat flux). The visualization, shown in Figure 8, is an animated version of figures from recently published results [59]. The visualization was highly effective in supporting the ocean scientists' investigation, leading to important research insights [68, 62], but it was limited to vertical and horizontal slices at predefined locations and required design, development, rendering, and distribution by the visualization researcher. Extracting vertical slices of high-resolution MITgcm data is computationally and I/O intensive due to the native layout of the simulation output. Restrictions on visualization and analysis prompted NASA visualization researchers to develop an HPC interactive vertical slicer, leveraging HPC resources (compute nodes, network, storage) at the NASA Advanced Supercomputing Division (NAS). Although an effective approach, the HPC vertical slicer requires dedicated compute nodes and dedicated time from visualization researchers. An interactive vertical slicer, accessible to and driven directly by ocean scientists, has been desired without having to be onsite at a supercomputer.

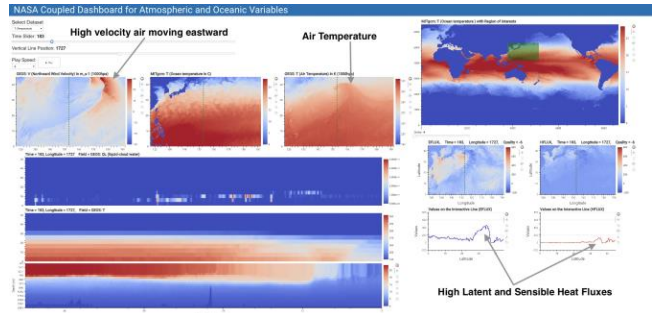

Figure 9: Increasing heat fluxes (two plots and images at the bottom right) and air temperature (image at top middle) create a high-velocity wind (image at top left) in the atmosphere moving eastward for the Kuroshio region.

Motivated both by the promising initial results and the limitations of the preliminary animation, the NASA/JPL/Caltech team began collaborating with a team from the Scientific Computing and Imaging Institute at the University of Utah to develop an interactive dashboard version of the preliminary visualization. The collaboration resulted in the dashboard shown in Figure 1 with the goal of helping study the relationships between different variables of ocean and atmospheric simulations at different regions of interest, such as the Gulf Stream region and Kuroshio region. For the Gulf Stream region, we use 75° W to 60° W and 30° N to 45° N. For the Kuroshio region, we use 117° E to 192° E and 0° N to 45° N. The dashboard leverages high-resolution datasets from the GEOS and MITgcm simulations to isolate and visualize the interplay between various climate variables. This region-specific approach allows scientists to observe how atmospheric conditions such as temperature and pressure gradients influence oceanic currents, salinity levels, and vertical velocities, and vice versa. Figure 10 shows how an investigation into an interesting wave pattern observed in the plots of sensible and latent heat flux (middle) and examination of the associated vertical slices led the ocean scientists to find that the wave pattern was trapped within the atmospheric boundary layer (left) and did not extend above the boundary layer, as previously thought.

### 5.3.3 Use Case 3: Data Democratization for Teaching

At Utah State Blanding, a Native American Serving Non-Tribal Institution, GEOG 4780/6780 Spatial Analysis is taught by Professor Gustavo Ovando-Montejo. This course, designed for upper-division undergraduates and graduates, has 25 students dedicating 4 to 6 hours weekly to spatial analysis using R. The curriculum emphasizes spatial reasoning, coding techniques, and GIScience tasks, including data manipulation, interpretation, and modeling, with a focus on spatial statistics like spatial regression.

Students in the course face significant challenges with data ac-quisition, particularly during their final projects, which account for 50% of their grades and involve selecting and analyzing their own data. To assist with this, we provided instructions for installing Jupyter Notebooks and accessing the LLC2160 dataset from the cloud, which includes large fields such as ocean velocities, temperature, and salinity. These datasets are massive, exceeding 1PB in total. Our Jupyter Notebook guides students through loading and visualizing these datasets, starting with an example of the salinity data, and allows them to select specific timesteps and regions for analysis. The first step in the Jupyter Notebook provides an example of loading the Salinity data field with dimensions of 8640*6480*90, with 10366 timesteps. The next step shows users how to load the data and select any timestep and region (x,y,z), at any quality or resolution they want. By Step 4 and within minutes of sitting down with the Jupyter Notebook, the students have a visualization of the NASA data in a plot. Additional steps in the notebook walk the students through querying the data, including calculating the percentage of voxels within the selected salinity range vs. calculating the percentage of world surface within the selected salinity range. The dashboard demonstration showed how students could seamlessly access spatial data stored in the cloud. Students were initially struck by the realization and excitement that they could access these global data and variables in full resolution, akin to methods used by NASA scientists.

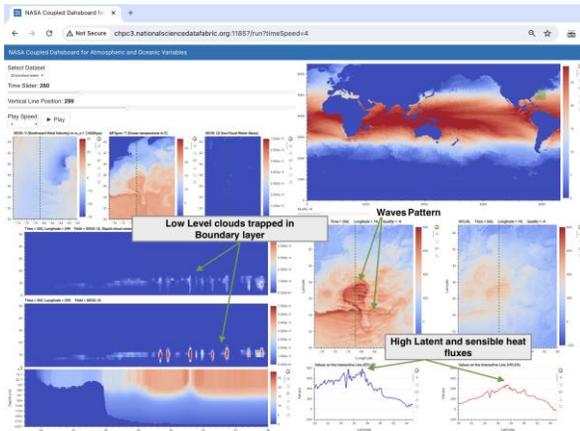

Figure 10: Interesting wave pattern observed in the plots of sensible and latent heat flux.

Students were excited to focus on specific areas of interest and visualize them in real-time. The dashboard allowed actual data analysis without downloading, though they could if desired. A highlight was a spatial query selecting voxels within a range, which students identified as basic suitability analysis — a notable achievement given the data's size. Overall, the dashboards were invaluable for hands-on GIScience teaching.

## 6 DISCUSSION

We report data conversion time and compression algorithm performance in terms of the peak signal-noise ratio for the DYAMOND and LLC4320 Ocean datasets. We also discuss the lessons learned as well as the impact on and benefits to society of our dashboards for petascale data democratization.

### 6.1 Performance for Climate Dashboards

We tested the dashboard objectively in terms of time to read the data at different locations and provide compression metrics.

**Time Comparison for data conversion and compression from different filesystems.** To test the efficiency of data conversion and compression techniques, we copied 48 timesteps for a 3D field of the same data, around 1 TB, to different locations, such as a personal computer and a server. We submitted a job to the Pleiades requesting one node and 24 cores to convert and compress the data.

It took around 1 hour and 20 minutes to convert and an additional 45 minutes to compress the data losslessly. The waiting queue for the job to run was around 10 minutes. The personal computer used for testing was a standard M1 Macbook with 16GB RAM and 8 cores. It took 1 hour 35 minutes to convert the same data and an additional 1 hour 20 minutes to compress it. The server we tested on had a 12-core x86-64 architecture Intel Xeon CPU with 64 GB RAM. Converting and compressing the same data as before took 50 minutes and 1 hour 20 minutes, respectively, on this server. Figure 11 shows these times in the same graph along with the total time for easier comparison.

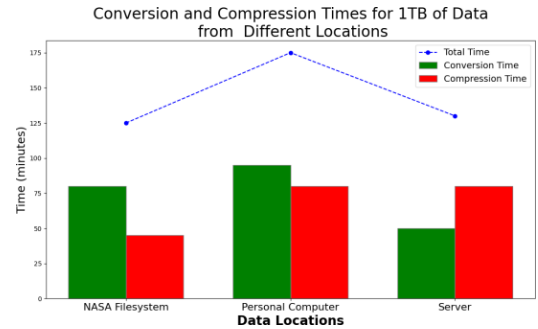

Figure 11: Result from an experiment done to estimate the time required to convert and compress 1 TB of data from different data locations, such as NASA Filesystem, personal computer, and a server.

**Performance Comparison.** Figure 12a analyzes data retrieval times from different storage systems at varying resolutions and locations. We compare NASA's filesystem, Pleiades [39], with cloud storage, both with and without local caching. Locally cached cloud data consistently provides quicker access than uncached data, with greater benefits as data resolution increases. This comparison highlights the effectiveness of local caching in optimizing access times, especially for high-resolution datasets, demonstrating significant advantages for exploring large-scale data.

**Data Quality Comparison.** After testing different lossy compression algorithms for the data, we present our use of *peak signal-to-noise ratio (PSNR)* as the metric for evaluating the data compression quality in Figure 12b. The PSNR is a measure used to assess the data reconstruction quality after lossy compressions. We compressed a sample of data using various precision bits with ZFP compression at varying data resolutions. The graph shows that higher bit precision, such as 32 and 30 bits, maintains a consistently high PSNR across all resolutions, indicating a minimal loss in data quality upon compression. On the other hand, lower precision levels, such as 24 and 16 bits, exhibit a considerable decrease in PSNR, suggesting a trade-off between compression level and data fidelity. The PSNR values remain fairly stable across different resolutions for each precision level.

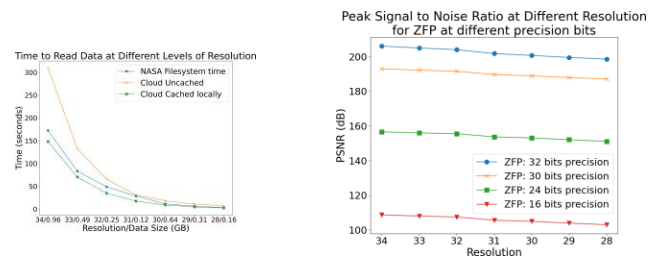

(a) Data retrieval time from different data locations at different levels of resolutions. (b) Peak signal-to-noise ratio for different bits of precision with ZFP compression algorithm.

Figure 12: Performance Analysis for NAS file systems, cloud uncached, and cloud cached (left) as well as the compression algorithms (right).

### 6.2 Lessons Learned

After several months of intermittent development and iteration, the NASA/JPL/Utah team met (half of the group met in person, half of the group joined remotely) to demonstrate and explore the results. During the session, the ocean scientists immediately demonstrated the ability to engage with the data in a way that was not possible before. Discussing phenomena of interest while interacting with the dashboard, they were able to develop new questions/hypotheses and provide preliminary answers about the interaction between the ocean and the atmosphere with a dramatic reduction of cognitive load that only interactive visualization can provide and has always been considered impossible for petascale data without supercomputing resources. Investigation into an interesting wave pattern observed in the plots of sensible and latent heat flux and examination of the associated vertical slices led the ocean scientists to find that the wave pattern was trapped within the atmospheric boundary layer and did not extend above the boundary layer, as previously thought. "This was something quite new for us because we thought it was much above the boundary layer, but no." The vertical slices in the dashboard shown on the left in Figure 1 show how the clouds are within the atmospheric boundary layer with very high and wavy latent and sensible heat fluxes.

Our collaboration with oceanic scientists showed the usefulness of faster visualizations and emphasized the need to be able to adjust timesteps and use a play button, which was crucial to seeing the expected correlation between multiple fields. Because each visualization generated in this interactive environment could take days to generate using the status quo fixed-frame animation approach, the domain scientists commented that our on-the-fly and interactive visualization "was astonishing." The ocean scientists also noted that the ability to interactively adjust the vertical slice location allowed them to check for numerical instabilities near vertical cuts, a capability they previously lacked. Additionally, the interactive dashboard lowered the barrier to exploration and collaboration by enabling investigations based on real-time data observations, facilitating more dynamic and immediate scientific inquiry.

Future iterations will help the new dashboards better match the traditional workflow of visualization and domain scientists. We need to be able to automate outputting animations for high-resolution data. However, as our use cases show, being able to set the resolution low for low bandwidth or quick investigations is something that climate scientists and visualization researchers could not easily do. The ability to share with college students the same data NASA and JPL scientists use for their research is creating an invaluable educational bridge. By enabling this accessibility, we're not only democratizing access to the latest scientific data but also inspiring the next generation of scientists and data analysts to explore the climate phenomena with the tools that leading researchers use.

### 6.3 Impact and Benefits

Traditional approaches to visualizing petascale simulation data rely on static animations with predefined variables and resolutions and/or supercomputing resources that are often combined with visualization expertise. The movement of massive datasets to the cloud, along with the progressive streaming capability, enables efficient access from platforms such as Jupyter Notebooks. This breakthrough benefits users without high-end computational resources and has significant practical implications in various fields.

**FAIR Data.** Our framework addresses the FAIR principles through the implementation of a FAIR Digital Object (FDO) framework, which democratizes access to large datasets. Our system ensures findability by including comprehensive metadata and identifiers within FDOs, making it easy to locate specific datasets. Accessibility is enhanced by providing an API that abstracts low-level storage intricacies, allowing users to request data without needing detailed knowledge of storage formats or network configurations. The framework supports interoperability by integrating multiple storage formats (e.g., ZARR, NetCDF, HDF5) and facilitating seamless data conversion, transfer, and caching across various platforms. Finally, reusability is promoted through the use of standardized metadata and executable elements within FDOs, ensuring that data remains useful and actionable for future research and applications.

**Remote Scientific Research and Collaboration.** The availability of massive datasets in a cloud-based, progressive streaming format significantly enhances the scope of remote scientific research and collaboration. Researchers from various parts of the world can now access complex datasets, such as those from the DYAMOND and MITgcm simulations, without the constraint of having powerful local computing resources. Our data fabric abstractions and dashboards facilitate a more inclusive and diverse research environment, enabling scientists to collaborate on global projects in areas such as climate modeling.

**Data Visualization and Analysis.** Our project's success opens up new possibilities in data visualization and analysis. Researchers and data analysts can now delve into complex datasets without worrying about the computational burdens traditionally associated with just visualizing the data. Broad audience accessibility allows groundbreaking insights and more effective communication of data-driven findings.

**Education and Training.** Educational institutions worldwide can benefit substantially from access to the NASA datasets. With a solution for the major problems of data storage and accessibility, students and educators can utilize these large-scale simulation data for learning and research purposes, enhancing the educational experience with massive data in data science and related fields.

### 7 CONCLUSION

In conclusion, we address several critical challenges inherent in the visualization and analysis of large-scale climate datasets, significantly impacting the scientific community. We introduce a novel data fabric abstraction layer that separates scientific needs from low-level specifications for file formats, network speed, and storage. We ensure efficient storage and data transfer speeds for petascale data via efficient data reorganization, conversion, and reduction/optimization pipeline with compression strategies and an Analysis-Ready Cloud-Optimized (ARCO) friendly format to significantly reduce the computational load and storage requirements. We also provide novel scalable visualization dashboards that enable progressive visualizations of petascale data with advanced analytical tools and a user-friendly design, encouraging scientific curiosity and discovery. Our framework results in data democratization for petascale data via publicly accessible web links to more than a petabyte of data on the cloud in an optimized format, enhancing public access and collaborative opportunities.

A rising tide of data democratization gives the most significant lift to boats that are usually the most excluded from access to the data. We encourage scientific curiosity in data by providing platform and device-agnostic dashboards for visualization and analysis platforms for streaming petascale data. Thus, a greater diversity of individuals or groups can access data, which benefits Science by backing up theories with scientific data or creating novel scientific theories based on captured data or generated simulations. Furthermore, driving greater inclusion does not require focusing on one specific group because any initiative that removes exclusionary practices can improve matters for all impacted groups. Our case studies are the boats catching the rising tide of data democratization, and we expect every boat full of data will rise as the tide of data equity raises us all.

### 8 GITHUB LINK FOR CODE AND SUPPLEMENTALS

https://github.com/sci-visus/Openvisus-NASA-Dashboard

**ACKNOWLEDGMENTS**

The authors wish to thank the NASA Ames Research Center, NASA Jet Propulsion Lab, Seal Storage, and ViSOAR LLC. NASA provided access to their data and computational resources via the Pleides supercomputer, and relevant use cases. Seal Storage provided cloud storage to host the data. ViSOAR LLC provided interface to dashboards via their secure server. This work was funded in part by NASA Cooperative Agreement 80NSSC23M0013, NASA JPL Subcontract No. 1685389 DOE SBIR Phase II #DE-SC0017152, NSF OAC award 2138811, NSF CI CoE Award 2127548, NSF OISE award 2330582, DoE award DE-FE0031880, the Intel oneAPI Centers of Excellence at University of Utah. Results presented in this paper were obtained in part using the Chameleon, Cloudlab, CloudBank, Fabric, and ACCESS testbeds supported by the National Science Foundation.

## REFERENCES

[1] R. Abernathey. Petabytes of Ocean Data, Part 1: NASA Ecco Data Portal. *Medium*, Jun 2019. 2

[2] R. P. Abernathey, T. Augspurger, A. Banihirwe, C. C. Blackmon-Luca, T. J. Crone, C. L. Gentemann, J. J. Hamman, N. Henderson, C. Lepore, T. A. McCaie, et al. Cloud-native repositories for big scientific data. *Computing in Science & Engineering*, 23(2):26–35, 2021. 2, 4

[3] H. Aizenman, M. Grossberg, D. Jones, N. Barnes, J. Smerdon, K. Anchukaitis, and J. E. Geay. Web Based Visualization Tool for Climate Data Using Python. In *92nd AMS Annual Meeting, Second Symposium on Advances in Modeling and Analysis Using Python. Sl: sn*, 2012. 2

[4] J. R. Alder and S. W. Hostetler. Web based visualization of large climate data sets. *Environmental Modelling & Software*, 68:175–180, 2015. 2

[5] E. Angel and E. Haines. An interactive introduction to WEBGL and three. JS. In *ACM SIGGRAPH 2017 Courses*, pp. 1–95. 2017. 2

[6] N. G. Author. *Synergistic Challenges in Data-Intensive Science and Exascale Computing. Summary report of the Advanced Scientific Computing Advisory Committee (ASCAC) Subcommittee, March 2013*. Mar. 2013. doi: 10.2172/1471113 3

[7] A. Barrett, C. Battisto, B. Bottomley, A. Friesz, A. Hunzinger, M. Jami, A. Lewandowski, B. Lind, L. Ló´pez, J. McNelis, and et al. NASA EarthData Cloud Cookbook. *Zenodo*, May 2023. doi: 10.5281/zenodo.7786710 2

[8] K. Begum, M. M. Rashid, and M. A. U. Shariff. Comparative Study of Big Data Visualization Tools and Techniques. In *Applied Informatics for Industry 4.0*, pp. 188–199. Chapman and Hall/CRC, 2023. 2

[9] H. Bhatia, D. Hoang, N. Morrical, V. Pascucci, P.-T. Bremer, and P. Lindstrom. AMM: Adaptive Multilinear Meshes. *IEEE Trans. on Visualization and Computer Graphics*, 28(6):2350–2363, 2022. 3

[10] Bokeh. https://bokeh.org/. Bokeh. 5

[11] M. Bostock, V. Ogievetsky, and J. Heer. D³ Data-Driven Documents. *IEEE Transactions on Visualization and Computer Graphics*, 17(12):2301–2309, 2011. doi: 10.1109/TVCG.2011.185 2

[12] N. R. Council. *Frontiers in Massive Data Analysis*. The National Academies Press, Washington, DC, 2013. doi: 10.17226/18374 3

[13] M. Cox and D. Ellsworth. Managing big data for scientific visualization. 01 1997. 1

[14] E. Daniel and F. Tschorsch. IPFS and friends: A qualitative comparison of next generation peer-to-peer data networks. *CoRR*, abs/2102.12737, 2021. 4

[15] D. A. Ellsworth, C. E. Henze, and B. C. Nelson. Interactive visualization of high-dimensional petascale ocean data. In *2017 IEEE 7th Symposium on Large Data Analysis and Visualization (LDAV)*, pp. 36–44, 2017. doi: 10.1109/LDAV.2017.8231849 2

[16] J. D. Ford, S. E. Tilleard, L. Berrang-Ford, M. Araos, R. Biesbroek, A. C. Lesnikowski, G. K. MacDonald, A. Hsu, C. Chen, and L. Bizikova. Big data has big potential for applications to climate change adaptation. *Proceedings of the National Academy of Sciences*, 113(39):10729–10732, 2016. 1

[17] A. Foussard, G. Lapeyre, and R. Plougonven. Response of Surface Wind Divergence to Mesoscale SST Anomalies under Different Wind Conditions. *J. of the Atmospheric Sciences*, 76:2065 – 2082, 2019. 7

[18] I. Fukumori, O. Wang, I. Fenty, G. Forget, P. Heimbach, and R. M. Ponte. ECCO Version 4 Release 3. Technical report, 2017. 2, 5

[19] C. L. Gentemann, C. A. Clayson, S. Brown, T. Lee, R. Parfitt, J. T. Farrar, M. Bourassa, P. J. Minnett, H. Seo, S. T. Gille, and V. Zlotnicki. FluxSat: Measuring the Ocean–Atmosphere Turbulent Exchange of Heat and Moisture from Space. *Remote Sensing*, 12(11), 2020. doi: 10.3390/rs12111796 7

[20] L. Grandinetti, G. R. Joubert, M. Kunze, and V. Pascucci, eds. *Big Data and High Performance Computing - Selected Papers from HPC Workshop, Cetraro, Italy, July 7-11, 2014*, vol. 26 of *Advances in Parallel Computing*. IOS Press, 2015. 3

[21] A. B. Gurvich and A. M. Geller. Firefly: A Browser-based Interactive 3D Data Visualization Tool for Millions of Data Points. *The Astrophysical Journal Supplement Series*, 265(2):38, mar 2023. 2

[22] D. Hoang, P. Klacansky, H. Bhatia, P.-T. Bremer, P. Lindstrom, and V. Pascucci. A Study of the Trade-off Between Reducing Precision and Reducing Resolution for Data Analysis and Visualization. *IEEE Transactions on Visualization and Computer Graphics*, 25(1):1193–1203, 2019. 3

[23] D. Hoang, A. Panta, G. Scorzelli, P. Davis, M. Parashar, and V. Pascucci. Publishing NASA's multi-petabytes of climate datasets, Dec 2022. 2

[24] D. Hoang, B. Summa, H. Bhatia, P. Lindstrom, P. Klacansky, W. Usher, P.-T. Bremer, and V. Pascucci. Efficient and Flexible Hierarchical Data Layouts for a unified encoding of scalar field precision and resolution. *IEEE Transactions on Visualization and Computer Graphics*, 27(2):603–613, 2021. 3

[25] Holoviz. Panel: The Powerful Data Exploration and Web App Framework for python. https://github.com/holoviz/panel. 5

[26] S. Hoyer and J. Hamman. XArray: N-D labeled Arrays and Datasets in Python. *Journal of Open Research Software*, 5(1):10, Apr. 2017. doi: 10.5334/jors.148 3

[27] A. Ismail, M. Toohey, Y. C. Lee, Z. Dong, and A. Y. Zomaya. Cost and Performance Analysis on Decentralized File Systems for Blockchain-Based Applications: State-of-the-Art Report. In *2022 IEEE International Conference on Blockchain (Blockchain)*, pp. 230–237, 2022. doi: 10.1109/Blockchain55522.2022.00039 4

[28] S. Jourdain, U. Ayachit, and B. Geveci. ParaViewWeb: A web framework for 3D visualization and data processing. *International Journal of Computer Information Systems and Industrial Management Applications*, 3(1):870–877, 2011. 2

[29] S. Klasky, J. Thayer, and H. Najm. *Data Reduction for Science: Brochure from the Advanced Scientific Computing Research Workshop*. Apr. 2021. doi: 10.2172/1770192 3

[30] H. Kotabe. Decentralized Storage: A Primer. https://www.tbstat.com/wp/uploads/2022/05/20220531_DecentralizedStorage_TheBlockResearch.pdf, May 2022. Accessed: 31/03/2024. 4

[31] S. Kumar, C. Christensen, J. A. Schmidt, P. Bremer, E. Brugger, V. Vishwanath, P. H. Carns, H. Kolla, R. W. Grout, J. Chen, M. Berzins, G. Scorzelli, and V. Pascucci. Fast Multiresolution Reads of Massive Simulation Datasets. In *29th International Supercomputing Conference, ISC 2014, Leipzig, Germany, June 22-26, 2014. Proceedings*, vol. 8488 of *Lecture Notes in Computer Science*, pp. 314–330. Springer, 2014. 3

[32] S. Kumar, S. Petruzza, W. Usher, and V. Pascucci. Spatially-Aware Parallel I/O for Particle Data. In *Proceedings of the 48th International Conference on Parallel Processing*, ICPP 2019. Association for Computing Machinery, New York, NY, USA, 2019. 3

[33] X. Li, Q. Liu, S. Wu, Z. Cao, and Q. Bai. Game theory based compatible incentive mechanism design for non-cryptocurrency blockchain systems. *J. of Industrial Information Integration*, 31:100426, 2023. 4

[34] P. Lindstrom. Fixed-rate compressed floating-point arrays. *IEEE transactions on visualization and computer graphics*, 20(12):2674–2683, 2014. 6

[35] X. Liu, X. Ma, P. Chang, Y. Jia, D. Fu, G. Xu, L. Wu, R. Saravanan, and C. M. Patricola. Ocean fronts and eddies force atmospheric rivers and heavy precipitation in western North America. *Nature Communications*, 12(1):1268, 2021. 7

[36] S. Lu, R. M. Li, W. C. Tjhi, K. K. Lee, L. Wang, X. Li, and D. Ma.

A Framework for Cloud-Based Large-Scale Data Analytics and Visualization: Case Study on Multiscale Climate Data. In *2011 IEEE Third International Conference on Cloud Computing Technology and Science*, pp. 618–622, 2011. doi: 10.1109/CloudCom.2011.95 2

[37] D. Menemenlis, C. Hill, C. Henze, J. Wang, and I. Fenty. Pre-SWOT Level-4 Hourly MITgcm LLC4320 Native 2km Grid Oceanographic Version 1.0, 2021. 5

[38] NASA. DYnamics of the Atmospheric general circulation Modelled On Non-hydrostatic Domains Phase. https://gmao.gsfc.nasa.gov/global_mesoscale/dyamond_phaseII/data_access/. 5

[39] NASA. NASA High-End Computing Capability. https://www.nas.nasa.gov/hecc/. 5, 6, 8

[40] NASA. SOTO by Worldview. https://soto.podaac.earthdatacloud.nasa.gov/. 2

[41] NASA. ECCO Data Portal, 2014. Accessed: Mar 29, 2024. 1

[42] NASA. GEOS/ECCO Coupled Nature Run Data Portal. https://data.nas.nasa.gov/geosecco/geoseccodata/c1440_llc2160/, 2021. Accessed: Mar 29, 2024. 1

[43] D. B. Olson and R. H. Evans. Rings of the Agulhas current. *Deep Sea Research Part A. Oceanographic Research Papers*, 33(1):27–42, 1986. 6

[44] OpenVisus. OpenVisus. https://github.com/sci-visus/openvisus. 3

[45] OpenVisus. OpenVisuspy. https://github.com/sci-visus/openvisuspy. 5

[46] V. Pascucci and R. J. Frank. Global static indexing for real-time exploration of very large regular grids. In G. Johnson, ed., *Proceedings of the 2001 ACM/IEEE conference on Supercomputing, Denver, CO, USA, November 10-16, 2001, CD-ROM*, p. 2. ACM, 2001. 3

[47] V. Pascucci, G. Scorzelli, B. Summa, P.-T. Bremer, A. Gyulassy, C. Christensen, S. Philip, and S. Kumar. The ViSUS visualization framework. In *High Performance Visualization*, pp. 439–452. Chapman and Hall/CRC, 2012. 3, 5

[48] C. Pernet, C. Svarer, R. Blair, J. D. V. Horn, and R. A. Poldrack. On the long-term archiving of research data. 2023. 4

[49] K. Potter, A. Wilson, P.-T. Bremer, D. Williams, C. Doutriaux, V. Pascucci, and C. Johhson. Visualization of uncertainty and ensemble data: Exploration of climate modeling and weather forecast data with integrated ViSUS-CDAT systems. *Journal of Physics: Conference Series*, 180:012089, July 2009. doi: 10.1088/1742-6596/180/1/012089 3

[50] R. Rajatheva. Performance Challenges with Data Visualizations in Browser Environment. Master's thesis, 2023. 2

[51] M. Raji, A. Hota, T. Hobson, and J. Huang. Scientific visualization as a microservice. *IEEE transactions on visualization and computer graphics*, 26(4):1760–1774, 2018. 2

[52] B. M. Randles, I. V. Pasquetto, M. S. Golshan, and C. L. Borgman. Using the Jupyter notebook as a tool for open science: An empirical study. In *2017 ACM/IEEE Joint Conference on Digital Libraries (JCDL)*, pp. 1–2. IEEE, 2017. 4

[53] Sci-Visus. Interactive visualization of petscale climate data using OpenVisus. https://github.com/sci-visus/Openvisus-NASA-Dashboard. 5

[54] Seal. Seal Storage. https://www.sealstorage.io/. 4

[55] J. Shen, Y. Li, Y. Zhou, and X. Wang. Understanding I/O performance of ipfs storage: a client's perspective. In *Proceedings of the International Symposium on Quality of Service*, IWQoS '19. Association for Computing Machinery, New York, NY, USA, 2019. 4

[56] C. Stern, R. Abernathey, J. Hamman, R. Wegener, C. Lepore, S. Harkins, and A. Merose. Pangeo forge: crowdsourcing analysis-ready, cloud optimized data production. *Frontiers in Climate*, 3:782909, 2022. 4

[57] B. Stevens, M. Satoh, L. Auger, J. Biercamp, C. S. Bretherton, X. Chen, P. Du¨ben, F. Judt, M. Khairoutdinov, D. Klocke, et al. DYAMOND: the DYnamics of the Atmospheric general circulation Modeled On Non-hydrostatic Domains. *Progress in Earth and Planetary Science*, 6(1):1–17, 2019. 5

[58] G. Strawn. Open Science, Business Analytics, and FAIR Digital Objects. In *2019 IEEE 43rd Annual Computer Software and Applications Conference (COMPSAC)*, vol. 2, pp. 658–663, 2019. 3

[59] E. Strobach, P. Klein, A. Molod, A. A. Fahad, A. Trayanov, D. Menemenlis, and H. Torres. Local Air-Sea Interactions at Ocean Mesoscale and Submesoscale in a Western Boundary Current. *Geophysical Research Letters*, 03 2022. doi: 10.1029/2021GL097003 7

[60] B. Summa, G. Scorzelli, M. Jiang, P.-T. Bremer, and V. Pascucci. Interactive editing of massive imagery made simple: Turning Atlanta into Atlantis. *ACM Trans. on Graphics (TOG)*, 30(2):1–13, 2011. 3

[61] X. Sun, S. Shen, G. G. Leptoukh, P. Wang, L. Di, and M. Lu. Development of a Web-based visualization platform for climate research using Google Earth. *Computers & Geosciences*, 47:160–168, 2012. 2

[62] H. Torres, P. Klein, L. Siegelman, F. Vivant, D. Menemenlis, A. Molod, E. Strobach, and N. McCurdy. Seasonality of air-sea coupling through submesoscale ocean gradients and latent heat fluxes in a Western Boundary Current. *In preparation for Geophysical Research Letters*, 2024. 7

[63] W. Usher, L. Dyken, and S. Kumar. Speculative Progressive Raycasting for Memory Constrained Isosurface Visualization of Massive Volumes. In *2023 IEEE 13th Symposium on Large Data Analysis and Visualization (LDAV)*, pp. 1–11, 2023. 2

[64] W. Usher, X. Huang, S. Petruzza, S. Kumar, S. R. Slattery, S. T. Reeve, F. Wang, C. R. Johnson, and V. Pascucci. Adaptive Spatially Aware I/O for Multiresolution Particle Data Layouts. In *2021 IEEE International Parallel and Distributed Processing Symposium (IPDPS)*, pp. 547–556, 2021. 3

[65] W. Usher and V. Pascucci. Interactive visualization of terascale data in the browser: Fact or fiction? In *2020 IEEE 10th Symposium on Large Data Analysis and Visualization (LDAV)*, pp. 27–36. IEEE, 2020. 2

[66] I. Vakilinia, W. Wang, and J. Xin. An Incentive-Compatible Mechanism for Decentralized Storage Network. *IEEE Transactions on Network Science and Engineering*, 10(4):2294–2306, July 2023. 4

[67] A. Venkat, C. Christensen, A. Gyulassy, B. Summa, F. Federer, A. Angelucci, and V. Pascucci. A Scalable Cyberinfrastructure for Interactive Visualization of Terascale Microscopy Data. *New York Scientific Data Summit (NYSDS) : proceedings*, 2016, 08 2016. 3

[68] F. Vivant, L. Siegelman, P. Klein, H. Torres, D. Menemenlis, A. Molod, E. Strobach, and N. McCurdy. Oceanic (sub)mesoscale fronts induce convective precipitations within atmospheric storms. *In preparation for Geophysical Research Letters*, 2024. 7

[69] J. D. Walker, B. H. Letcher, K. D. Rodgers, C. C. Muhlfeld, and V. S. D'Angelo. An Interactive Data Visualization Framework for Exploring Geospatial Environmental Datasets and Model Predictions. *Water*, 12(10), 2020. doi: 10.3390/w12102928 2

[70] Y. Wang. Deck.gl: Large-scale web-based visual analytics made easy. *arXiv preprint arXiv:1910.08865*, 2019. 2

[71] M. D. Wilkinson, M. Dumontier, I. J. Aalbersberg, G. Appleton, M. Axton, A. Baak, N. Blomberg, J.-W. Boiten, L. B. da Silva Santos, P. E. Bourne, et al. The FAIR Guiding Principles for scientific data management and stewardship. *Scientific data*, 3(1):1–9, 2016. 3

[72] D. N. Williams, T. Bremer, C. Doutriaux, J. Patchett, S. Williams, G. Shipman, R. Miller, D. R. Pugmire, B. Smith, C. Steed, E. W. Bethel, H. Childs, H. Krishnan, P. Prabhat, M. Wehner, C. T. Silva, E. Santos, D. Koop, T. Ellqvist, J. Poco, B. Geveci, A. Chaudhary, A. Bauer, A. Pletzer, D. Kindig, G. L. Potter, and T. P. Maxwell. Ultrascale Visualization of Climate Data. *Computer*, 46(9):68–76, 2013. doi: 10.1109/MC.2013.119 2

[73] P. C. Wong, H.-W. Shen, and C. Chen. Top ten interaction challenges in extreme-scale visual analytics. *Expanding the frontiers of visual analytics and visualization*, pp. 197–207, 2012. 2

[74] P. C. Wong, H.-W. Shen, R. Leung, S. Hagos, T.-Y. Lee, X. Tong, and K. Lu. Visual analytics of large-scale climate model data. In *2014 IEEE 4th Symposium on Large Data Analysis and Visualization (LDAV)*, pp. 85–92, 2014. doi: 10.1109/LDAV.2014.7013208 2

[75] xmitgcm. Xmitgcm. https://xmitgcm.readthedocs.io/en/latest/. 2

[76] S.-E. Yoon, P. Lindstrom, V. Pascucci, and D. Manocha. Cache-oblivious mesh layouts. *ACM Transactions on graphics: ACM SIGGRAPH 2005 Papers*, 24(3):886–893, 2005. 3

[77] L. Zepner, P. Karrasch, F. Wiemann, and L. Bernard. ClimateCharts.net–an interactive climate analysis web platform. *International Journal of Digital Earth*, 14(3):338–356, 2021. 2